# 'Exponential' Quantum Tsallis-Havrda- Charvat Entropy of 'Type $\alpha$'

Dhanesh Garg, Satish Kumar

Maharishi Markandeshwar University

Mullana, Ambala-133001(Haryana), India

November 4, 2015

## Abstract

Entropy is a key measure in studies related to information theory and its many applications. Campbell of the first time recognized that exponential of Shannon's entropy is just the size of the sample space when the distribution is uniform. In this paper, we introduce a quantity which is called exponential Tsallis-Havrda-Charvat entropy and discuss its some properties. Further, we gave the application of exponential Tsallis-Havrda-Charvat entropy in quantum information theory which is called exponential quantum Tsallis-Havrda-Charvat entropy with its some major properties such as non-negative, concavity and continuity. It is found that projective measurement will not decrease the quantum entropy of a quantum state and at the end of the paper gave an upper bound on the quantum exponential entropy in terms of ensembles of pure state.

## Keywords

Convex Function, Exponential Shannon's entropy, Exponential Tsalli's entropy, Von-Neumann entropy, Exponential quantum Tsalli's entropy.

## 1. Introduction

Let $A = (a_1, a_2, ..., a_n) \in \Delta_n$, where $\Delta_n = \left\{ (a_1, a_2, ..., a_n),\ a_i \geq 0,\ i = 1, 2, ..., n,\ n \geq 2,\ \sum_{i=1}^{n} a_i = 1 \right\}$ be a set of discrete finite n-ary probability distributions. Havrda and Charvat [4] defined an entropy of degree $\alpha$ as

$$^{*}H^{\alpha}(A) = \frac{1}{1 - 2^{1-\alpha}} \left[ 1 - \sum_{i=1}^{n} a_i^{\alpha} \right], \tag{1}$$

where $\alpha$ is a real positive parameter not equal to one. Independently Tsallis [14], proposed a one parameter generalization of the Shannon entropy as

$$H^{\alpha}(A) = \frac{1}{\alpha - 1} \left[ 1 - \sum_{i=1}^{n} a_i^{\alpha} \right], \tag{2}$$

where $\alpha$ is a real positive parameter and $\alpha \neq 1$. Both these entropies essentially have the same expression except the normalized factor. The Havrda and Charvat entropy is normalized to 1. That is, if $A = \left(\frac{1}{2}, \frac{1}{2}\right)$, then $^*H^\alpha(A) = 1$, where as the Tsallis entropy is not normalized. Both the entropies yield the same result and we called these entropies as the Tsallis-Havrda-Charvat entropy. Since $\lim_{\alpha \to 1} H^\alpha(A) = H(A)$, the Tsallis-Havrda-Charvat entropy $H^\alpha(A)$ is a single parameter generalization of the Shannon entropy $H(A)$, where

$$H(A) = -\sum_{i=1}^{n} a_i \log a_i. \qquad (3)$$

When $\alpha = 2$, the Tsallis-Havrda-Charvat entropy becomes the Gini-Simpson index of diversity,

$$H^2(A) = \left[1 - \sum_{i=1}^{n} a_i^2\right]. \qquad (4)$$

Tsallis-Havrda-Charvat entropy was studied as the entropy of degree $\alpha$ for the first time by Havrda and Charvat [4] and then by Daroczy [3]. This entropy become often used in statistical physics after the seminal work of Tsallis [14]. Various generalized entropies have been introduced in the literature, taking the Shannon entropy as basic and have found applications in various disciplines such as economics, statistics, information processing and computing etc. Generalizations of Shannon's entropy started with Renyi's entropy [12] of order- $\alpha$, given by

$$H_\alpha(A) = \frac{1}{1-\alpha} \log \left[\sum_{i=1}^{n} a_i^\alpha\right], \quad \alpha \neq 1, \alpha > 0 \qquad (5)$$

Campbell [2] studied exponentials of the Shannon's and Renyi's entropies, given by

$$E(A) = e^{H(A)} \qquad (6)$$

and

$$E_\alpha(A) = e^{H_\alpha(A)}, \qquad (7)$$

where $H(A)$ and $H_\alpha(A)$ represent respectively the Shannon's and Renyi's entropies. It may also be mentioned that Koski and Persson [8] studied

$$E_{(\alpha,\beta)}(A) = e^{H_{(\alpha,\beta)}(A)}, \qquad (8)$$

exponential of Kapur's entropy [7] given by

$$H_{(\alpha,\beta)}(A) = \frac{1}{(\beta - \alpha)} \log \frac{\sum_{i=1}^{n} a_i^\alpha}{\sum_{i=1}^{n} a_i^\beta}, \quad \alpha \neq \beta, \alpha, \beta > 0 \qquad (9)$$

It is interesting to notice that, in the case of discrete uniform distribution $A \in \Delta_n$, (6),(7), and (8) all reduce to $n$, just the 'size of the sample space of the distribution'.

This paper is organized as follows: Sec. II, define the exponential Tsallis-Havrda-Charvat entropy and discuss its some major properties corresponding to Tsallis-Havrda-Charvat entropy. Sec. III, define exponential quantum Tsallis entropy and some important properties such as non-negative, continuity, concavity are proved but different from Von Neumann entropy. Sec. IV, it is found that the projective measurement will not decrease the exponential quantum entropy of a quantum state and gave an upper bound on the exponential quantum Tsallis entropy in terms of ensembles of pure state.

In the next section, we study some properties of the exponential 'type $\alpha$' entropy.

## 2. Exponential 'Type $\alpha$' Entropy and Its Properties

Corresponding to Tsallis-Havrda-Charvat 'type $\alpha$' entropy, the exponential 'type $\alpha$' entropy is defined as follows:

**Definition:** Exponential 'type $\alpha$' entropy of a discrete distribution A is given by:

$$E^\alpha(A) = \frac{1 - e^{\left[\sum_{i=1}^n p_i^\alpha\right] - 1}}{\alpha - 1}, \quad \alpha > 0 (\neq 1). \tag{10}$$

When $\alpha \to 1$, measure (10) reduces to Shannon's entropy. The quantity (10) is entropy. Such a name will be justified, if it shares some major properties with Shannon's and other entropies in the literature. We study some such properties in the following theorem.

**Theorem 2.1:** The measure of information $E^\alpha(A)$, $A = (a_1, a_2, ..., a_n) \in \Delta_n$, where $\Delta_n = \{(a_1, a_2, ..., a_n) : a_i \geq 0, \sum_{i=1}^n a_i = 1\}$ has the following properties:

1. Symmetry: $E^\alpha(A) = E^\alpha(a_1, a_2, ..., a_n)$ is a symmetric function of $(a_1, a_2, ..., a_n)$.

2. Non-negative: $E^\alpha(A) \geq 0$, for all $\alpha > 0 (\neq 1)$.

3. Expansible:
$$E^\alpha(a_1, a_2, ..., a_n; 0) = E^\alpha(a_1, a_2, ..., a_n).$$

4. Decisive:
$$E^\alpha(0, 1) = E^\alpha(1, 0) = 0.$$

5. Maximality:
$$E^\alpha(a_1, a_2, ..., a_n) \leq E^\alpha(1/n, 1/n, ..., 1/n) = \frac{1}{\alpha - 1}\left[1 - e^{\left[n^{1-\alpha} - 1\right]}\right].$$

6. Concavity: The measure $E^\alpha(A)$ is a concave function of the probability distribution $A = (a_1, a_2, ..., a_n), a_i \geq 0, \sum_{i=1}^n a_i = 1$, for all $\alpha > 0 (\neq 1)$.

7. Continuity: $E^\alpha(a_1, a_2, ..., a_n)$ is continuous in the region $a_i \geq 0$ for all $\alpha > 0$.

**Proof:** (1),(3),(4) and (5): these properties are obvious and can be and verified easily. We know that $e^{\left[\sum_{i=1}^n p_i^\alpha\right] - 1} - 1$ is continuous in the region $a_i \geq 0$ for all $\alpha > 0$. Hence, $E^\alpha(A)$, is also continuous in the region $a_i \geq 0$ for all $\alpha > 0$.

**Property 2:** The measure $E^\alpha(A)$ is non-negative for all $\alpha > 0 (\neq 1)$.

**Proof:** We consider the following cases:

**Case (i):** When $\alpha > 1$,
$$\Rightarrow \quad \left[1 - e^{\left[\sum_{i=1}^n p_i^\alpha\right] - 1}\right] > 0,$$

and $\alpha - 1 > 0$, we get

$$E^\alpha(A) > 0. \tag{11}$$

**Case (ii):** When $0 < \alpha < 1$,

$$\Rightarrow \quad \left[1 - e^{[\sum_{i=1}^n p_i^\alpha] - 1}\right] < 0,$$

and $\alpha - 1 < 0$, we get

$$E^\alpha(A) > 0. \tag{12}$$

From (11) and (12), we conclude that $E^\alpha(P) > 0$ for all $\alpha > 0$.

To prove the next property, we shall use the following definition of a concave function.

**Definition 4 (Concave Function):** A function $f(.)$ over the points in a convex set $\mathbb{R}$ is *concave* if for all $r_1, r_2 \in \mathbb{R}$ and $\mu \in (0,1)$, then

$$\mu f(r_1) + (1-\mu) f(r_2) \leq f(\mu r_1 + (1-\mu) r_2) \tag{13}$$

The function $f(.)$ is convex if the above inequality holds with $\geq$ in place of $\leq$.

**Property 6:** The measure $E^\alpha(A)$ for all $\alpha > 0 \, (\neq 1)$ is a concave function of the probability distribution $A = (a_1, ..., a_n)$, $a_i \geq 0$, $\sum_{i=1}^n a_i = 1$.

**Proof:** Associated with the random variable $X = (x_1, x_2, ..., x_n)$, let us consider $r$ distributions

$$A_k(X) = \{a_k(x_1), a_k(x_2), ..., a_k(x_n)\},$$

where

$$\sum_{i=1}^n a_k(x_i) = 1, \; k = 1, 2, ..., r.$$

Next let there be $r$ numbers $(\lambda_1, \lambda_2, ..., \lambda_r)$ such that $\lambda_k \geq 0$, $\sum_{k=1}^r \lambda_k = 1$ and define

$$A_0(X) = \{a_0(x_1), a_0(x_2), ..., a_0(x_n)\},$$

where

$$a_0(x_i) = \sum_{k=1}^r \lambda_k p_k(x_i), \; i = 1, 2, ..., n.$$

Obviously $\sum_{i=1}^n a_0(x_i) = 1$, and thus $A_0(X)$ is a bonafide distribution of $X$.

If $\alpha > 1$, then we have

$$\sum_{k=1}^r \lambda_k E^\alpha(A_k) - E^\alpha(A_0)$$

$$= \sum_{k=1}^r \lambda_k E^\alpha(A_k) - \frac{\left[1 - e^{\left(\sum_{i=1}^r \lambda_i p_i\right)^\alpha - 1}\right]}{\alpha - 1}$$

$$\leq \sum_{k=1}^r \lambda_k E^\alpha(A_k) - \frac{\left[1 - e^{\left(\sum_{i=1}^r a_i p_i^\alpha\right) - 1}\right]}{\alpha - 1}$$

i.e.,

$$\sum_{i=1}^r \lambda_i E^\alpha(A_i) \leq E^\alpha(A) \tag{14}$$

Similarly for $0 < \alpha \leq 1$, (14) also holds.

In the next section, we will give the applications of exponential Tsallis entropy in quantum information theory.

## 3. Exponential Quantum Entropy and its Properies

The Von Neumann entropy is the quantum version of Shannon entropy. Von Neumann defined the entropy of a quantum state $\rho$ by the formula

$$S(\rho) = -tr\left[(\rho)\log(\rho)\right], \tag{15}$$

where $\rho$ is any density operator, i.e., positive operator on a complex separable Hilbert space $H$ having a unit trace. The exponential Tsallis-Havrda-Charvat entropy also have their quantum version, we called it exponential quantum Tsallis-Havrda-Charvat entropy and is defined as:

$$E^\alpha(\rho) = \frac{1}{\alpha - 1}\left[1 - e^{[tr(\rho^\alpha)]-1}\right], \alpha > (\neq 1). \tag{16}$$

Using the spectral decomposition theorem and noting $tr(U\rho U^\dagger) = tr(\rho)$, where U is unitary, we can easily verify that

$$\lim_{\alpha \to 1} E^\alpha(\rho) = S(\rho) \tag{17}$$

The quantity (16) is exponential quantum entropy such a name will be justified, if it shares some majors properties with the other's properties in the literature such as non-negativity, concavity, continuity etc.

Now we will present some properties in the following proposition.

**Proposition 1:** The exponential quantum entropy is non-negative, i.e., $E_\alpha(\rho) \geq 0$, where $\rho$ is any density operator. Then $E_\alpha(\rho) = 0$, if and only if the state is pure.

**Proof:** The Von Neumann entropy $S(\rho) \geq 0$, [Ref. 9 &15]. Now we prove $E_\alpha(\rho) \geq 0$. One has $x^\alpha \leq (resp. \geq 0)x$ for $x \in [0,1]$ with equality iff x=0 or x=1 when $\alpha > 1(resp. 0 < \alpha < 1)$. Since the eigenvalues $p_j$ of $\rho$ lie in [0, 1], we conclude the operator inequality $\rho^\alpha \leq 1(\ resp. \ \geq)\rho$, where equality holds iff $\rho$ is pure, that is, a one-dimensional projection. It then follows that $e^{tr(\rho^\alpha)-1} \leq (resp.) \geq 1$ with equality iff $\rho$ is pure. Noting the sign of $\frac{1}{\alpha-1}$ in different cases, we complete the proof.

**Proposition 2:** In a finite dimension $n, E^\alpha(\rho)$ is bounded, i.e.,

$$E^\alpha(\rho) \leq \begin{cases} \log{(rank\,\rho)} & ; if\,\alpha = 1 \\ \frac{1}{\alpha-1}\left[1 - e^{(rank\rho)^{1-\alpha}-1}\right] & ; if\,\alpha > 0 (\neq 1) \end{cases}. \tag{18}$$

Equality holds in the inequality if and only if $\rho$ is an equidistribution of order $rank\,(\rho)$.

**Proof:** To prove the upper bounds of (18), we use Holder's inequality. For $0 < \alpha < 1$,

$$tr(\rho^\alpha) = \sum\nolimits_{j=1}^{rank\,\rho} p_j^\alpha \leq \left(\sum\nolimits_{j=1}^{rank\,\rho}(p_j^\alpha)^{\frac{1}{\alpha}}\right)^\alpha \left(\sum\nolimits_{j=1}^{rank\,\rho} 1^{\frac{1}{1-\alpha}}\right)^{1-\alpha} = (rank\,\rho)^{1-\alpha}, \tag{19}$$

This implies $e^{[tr(\rho^\alpha)]-1} \leq e^{\left[(rank\rho)^{1-\alpha}\right]-1}$,

where $p_j$ is the eigenvalues of $\rho$.

Equality holds iff $p_j$ is constants for $j = 1, 2, ..., rank(\rho)$, i.e., $p_j = (rank(\rho))^{-1}$.

We have the result as follows:

$$E^\alpha(\rho) \leq \frac{1}{1-\alpha}\left[1 - e^{(rank(\rho))^{1-\alpha}-1}\right], \alpha > 0 (\neq 1). \tag{20}$$

Which is the second inequality in (18). For $\alpha > 1$, the inequality in (19) is reversed but the conditions for equality are the same.

For $\alpha = 1$, Von Neumann entropy is at most $\log(rank\,\rho)$. This completes the proof.

Concavity is another important property. $E^\alpha(\rho)$ similar to the Von Neumann entropy. To prove the concavity property we have the following lemma.

**Lemma 1:** Let $G_\alpha(\rho) = tr(\rho^\alpha)$, $\alpha > 0 (\neq 1)$, then

1. $G_\alpha(\rho)$ is a concave function of the density operator $\rho$ for $0 < \alpha < 1$.

2. $G_\alpha(\rho)$ is a convex function of the density operator $\rho$ for $\alpha \geq 1$.

**Proof:** We apply Minkowski's inequality to positive operators: Let $\rho, \sigma$ be positive operators, and $1 < \alpha < \infty$, then

$$[tr(\rho+\sigma)^\alpha]^{\frac{1}{\alpha}} \leq [tr(\rho)^\alpha]^{\frac{1}{\alpha}} + [tr(\sigma)^\alpha]^{\frac{1}{\alpha}}. \tag{21}$$

The inequality (21) is reversed if $0 < \alpha < 1$

**Case 1:** When $0 < \alpha < 1$, from (21), we have

$$[tr(\lambda\rho + \mu\sigma)^\alpha] \geq \left[\lambda[tr(\rho)^\alpha]^{\frac{1}{\alpha}} + \mu[tr(\sigma)^\alpha]^{\frac{1}{\alpha}}\right]^\alpha. \tag{22}$$

If $0 < \alpha < 1$, the function $y = x^\alpha$ is concave. This gives

$$\left[\lambda[tr(\rho)^\alpha]^{\frac{1}{\alpha}} + \mu[tr(\sigma)^\alpha]^{\frac{1}{\alpha}}\right]^\alpha \geq \lambda[tr(\rho)^\alpha] + \mu[tr(\sigma)^\alpha]. \tag{23}$$

From (22) and (23), we get

$$tr\,(\lambda\rho + \mu\sigma)^\alpha \geq \lambda[tr(\rho)^\alpha] + \mu[tr(\sigma)^\alpha]. \tag{24}$$

Where $\lambda, \mu \geq 0$, $\lambda + \mu = 1$. Therefore $G_\alpha(\rho)$ is a concave function for $0 < \alpha < 1$.

**Case 2:** When $\alpha \geq 1$, the inequalities in (24) is reversed,

$$tr\,(\lambda\rho + \mu\sigma)^\alpha \leq \lambda[tr(\rho)^\alpha] + \mu[tr(\sigma)^\alpha] \tag{25}$$

Where $\lambda, \mu \geq 0$, $\lambda + \mu = 1$. Therefore $G_\alpha(\rho)$ is a convex function for $\alpha \geq 1$.

Concavity of $E^\alpha(\rho)$ follows from Lemma 1 and the sign of $\frac{1}{\alpha-1}$.

**Proposition 3:** Suppose $\rho$ and $\sigma$ are density operators and $0 \leq \lambda \leq 1$. Then
$E^\alpha(\lambda \rho + (1-\lambda)\sigma) \geq \lambda E^\alpha(\rho) + (1-\lambda)E^\alpha(\sigma)$ for all $\alpha$ satisfying $\alpha > 0 (\neq 1)$.

**Remark:** For $\alpha = 1$, the Von Neumann entropy is a concave function.

### 4. Projective Measurement and Quantum Exponential Entropy

In this section, it is shown that projective measurement will not decrease the exponential quantum Tsallis-Havrda-Charvat entropy of a quantum state, either. And we give an upper bound on the exponential quantum entropy (16) in terms of ensembles of pure states. How does the exponential entropy of a quantum system behave when we perform a measurement on that system? Not surprisingly, the answer to this question depends on the type of measurement which we perform. Nevertheless, there are some surprisingly general assertions we can make about how the entropy behaves.

Suppose for example, than an orthogonal measurement described by projectors $P_i$ is performed on a quantum system but we never learn the result of the measurement. If the state of the system before the measurement was $\rho$, then the state after is given by

$$\rho' = \sum_{i=1}^{n} P_i \rho P_i. \tag{26}$$

The following theorem shows that the exponential entropy is never decreased in this case and remains the same only when the state is not changed by the measurement.

**Theorem 4.1:( Projective measurements will not decrease the Exponential entropy):**
Suppose $P_i$ is complete set of orthogonal projectors and $\rho$ is a density operator. Then the entropy of the state $\rho' = \sum_{i=1}^{n} P_i \rho P_i$ of the system after measurement is at least as greater as the original exponential entropy,

$$E_\alpha(\rho') \geq E^\alpha(P), \tag{27}$$

with the equality iff $\rho = \rho'$, where $E^\alpha(\rho)$ represent the quantum exponential entropy of 'type $\alpha$' and it is defined as in (16).

**Proof:** Spectral decomposition of state $\rho$ can be expressed as

$$\rho = \sum_a P(a)|a><a| \tag{28}$$

Where $P(a) \in [0,1]$, $\sum_a P(a) = 1$, and $\{|a>\}$ represents a complete orthogonal base. Based on this, we can express state $\rho'$ as

$$\rho' = \prod(P) = \sum_i P_i \left[\sum_a P(a)|a><a|\right] P_i. \tag{29}$$

Writing $P_i = |i><i|$ with $<i|i> = \delta_{ii'}$, we have ($\{|i>\}$ represents a complete orthogonal base)

$$\prod(|a><a|) = \sum_i |<a|i>|^2 |i><i|. \tag{30}$$

Thus we find

$$(\rho')^\alpha = [\prod(P)]^\alpha = \sum_i \left[\sum_a P(a)|<a|i>|^2\right]^\alpha |i><i|, \qquad (31)$$

which can lead to

$$tr\left[(\rho')^\alpha\right] = \sum_i \left[\sum_a P(a)|<a|i>|^2\right]^\alpha. \qquad (32)$$

Here, we can note [6] that

$$\sum_a |<a|i>|^2 = \sum_i |<a|i>|^2 = 1. \qquad (33)$$

Since $\alpha > 1$, in that case we can note that function $f(x) = x^\alpha (x > 0)$ is convex.

So,

$$f\left[\sum_i \lambda_i a_i\right] \leq \sum_i \lambda_i f(a_i), \qquad (34)$$

where $\lambda_i \in (0,1)$ and $\sum_i \lambda_i = 1$. From this, (30) and (31), it follows

$$tr\left[(\rho')^\alpha\right] = \sum_i \left[\sum_a P(a)|<a|i>|^2\right]^\alpha.$$
$$\leq \sum_a [P(a)]^\alpha \sum_i |<a|i>|^2$$
$$= \sum_a [P(a)]^\alpha$$
$$\Rightarrow tr[(\rho')] \leq \sum_a [P(a)]^\alpha.$$
$$\Rightarrow e^{[tr[(\rho')]]-1} \leq e^{\sum_a [P(a)]^\alpha - 1}. \qquad (35)$$

After simplification of (35), we get

$$E_\alpha(\rho') \geq E^\alpha(P),$$

which means, that projective measurement can only increase the exponential entropy of the state for $\alpha > 1$. Similarly for the case $0 < \alpha < 1$, but the inequality in (35) is reversed.

Now we obtain a relation between the exponential entropy and quantum exponential entropy.

**Theorem 4.2:** Let $\{p_i|\psi_i>\}$, where $<\psi_i|\psi_i>=1$ and $\sum_i p_i = 1$, be an ensemble of pure states that gives rise to the density operator $\rho$, i.e.,

$$\rho = \sum_i p_i |\psi_i><\psi_i|. \qquad (36)$$

For $\alpha > 1$, there holds

$$E^\alpha(\rho) \leq E^\alpha(P). \qquad (37)$$

**Proof:** Let $\lambda_j$ and $|\phi_j>$ denote eigenvalues and eigen states of $\rho = \sum_j \lambda_j |\phi_j><\phi_j|$. The ensemble classification theorem say that [5]

$$\sqrt{p_i}|\psi_i> = \sum_j u_{ij}\sqrt{\lambda_j}|\phi_j> \qquad (38)$$

for some unitary matrix $[[u_{ij}]]$. It follows from (38) and $<\phi_j|\phi_k>=\delta_{jk}$ that $p_i = \sum_j w_{ij}\lambda_j$, where $w_{ij} = u_{ij}^* u_{ij}$ are elements of a unistochastic matrix, i.e., $\sum_i w_{ij} = 1$ for all $j$ and $\sum_j w_{ij} = 1$ for all $i$. The function $f(x) = x^\alpha$ $(x > 0)$ is convex for $\alpha > 1$. Applying Jensen's inequality to this function, we obtain

$$\sum_i p_i^\alpha = \sum_i \left[\sum_j w_{ij} \lambda_j\right]^\alpha \leq \sum_i \sum_j w_{ij} \lambda_j^\alpha = \sum_j \lambda_j^\alpha,$$

in view of uni-stochasticity of the matrix $[[w_{ij}]]$.

$$\sum_i p_i^\alpha \leq \sum_j \lambda_j^\alpha. \tag{39}$$

Since the factor $\frac{1}{\alpha-1} > 0$ for $\alpha > 1$ and we get,

$$\frac{1}{\alpha-1}\left[1 - e^{(\sum_i p_i^\alpha)-1}\right] \geq \frac{1}{\alpha-1}\left[1 - e^{(\sum_j \lambda_j^\alpha)-1}\right]. \tag{40}$$

Due to $\sum_j \lambda_j^\alpha = tr(p^\alpha)$ and the definition (16), the inequality (40) provides (31).

The statement of the Theorem 4.2 gives an upper bound on the exponential quantum Tsallis-Havrda-Charvat entropy in terms of ensembles of pure state. So the scope of the inequality (31) is much wider for the parameter $\alpha$.

## 5. References

1. Arndt, C., Information Measure-Information and its description in Science and Engineering, Springer, Berlin, (2001).

2. Campbell, L. L., Exponential entropy as a measure of extent of distribution, Z. Wahrscheinlichkeitstheorie verw. Geb, vol. 5, pp. 217- 225,1966.

3. Daroczy, Z., "Generalized Information function," Information and Control, Vol. 16, pp. 36-51 (1970).

4. Havrda, J.F. and Charvat, F., "Qualification Method of Classification Process, the concept of structural $\alpha$-entropy," Kybernetika, Vol.3, pp. 30-35 , 279 (1967).

5. Hugkston, L.P., Jozsa, R. and Wootters, W.K., "A complete classification of quantum ensembles having a given density matrix,"Phys. Lett. A., Vol. 183, pp. 14-18, (1993).

6. Jankovic, M.V., Quantum Tsallis entropy and Projective Measurement, arxiv preprint arxiv: 0904.3794,2009-arxiv.org.physics (2009).

7. Kapur,J.N., Measure of information and their applications,1st ed., New Delhi, Wiley Eastern Limited, 1994.

8. Koski, T. and L. E. Persson, Some properties of generalized exponential entropies with applications to data compression, Information Sciences, vol. 62, pp. 103-132, 1992.

9. Nielsen, M.A. and Chuang, I.L.' "Quantum Computation and Quantum Information," Cambridge University Press, Cambridge, (2000).

10. Nielsen, F. and Nock, R., On R'enyi and Tsallis entropies and divergences for exponential families, 2011. http://arxiv.org/abs/1105.3259v1.

11. Petz, D.' Quantum information theory and Quantum statistics, Springer, Berlin,(2008).

12. Renyi, A. , On measures of entropy and information, in proceeding of the Forth Berkeley Symposium on Mathematics, Statistics and Probability- 1961, pp. 547-561, 1961.

13. Shannon, C.E., A Mathematical Theory of Communication, Bell System Tech. J., Vol.27, 379-423, 623-656(1948).

14. Tsallis, C., Possible generalization of Boltzmann–Gibbs statistics,J. Stat. Phys. Vol. 52, 480–487(1988).

15. Wehrl, A., "General properties of entropy," Rev. Mod. Phys., Vol. 50(**??**), pp. 221-260, (1978).